\documentclass{article}

\usepackage[english]{babel}
\usepackage[utf8]{inputenc}
\usepackage{amsmath}
\usepackage{amsfonts}
\usepackage{graphicx}
\usepackage[colorinlistoftodos]{todonotes}
\usepackage{blindtext}
\usepackage{amssymb}
\usepackage{amsthm}
\usepackage[toc]{appendix}
\usepackage{algorithm}

\newcommand{\textR}[1]{{\tt #1}}

\theoremstyle{remark}

\theoremstyle{definition}

\title{Survival analysis as a classification problem}

\author{Chenyang Zhong and Robert Tibshirani\\
Departments of Biomedical Data Science and Statistics,\\ Stanford University } 

\date{\today}

\begin{document}

\maketitle
\begin{abstract}
    In this paper, we explore a method for treating survival analysis as a classification problem. The method uses a ``stacking'' idea that collects the features and outcomes of the survival data in a large data frame, and  then treats it as a classification problem. In this framework, various statistical learning algorithms (including logistic regression, random forests, gradient boosting machines and neural networks) can be applied to estimate the parameters and make predictions. For stacking with logistic regression, we show that this approach is approximately equivalent to the Cox proportional hazards model with both theoretical analysis and simulation studies. For stacking with other machine learning algorithms, we show through simulation studies that our method  can outperform Cox proportional hazards model in terms of estimated survival curves. This idea is not new, but we believe that it should be better known by statistiicians
    and other data scientists.

\end{abstract}
\section{Introduction}

\subsection{Basic Problem}
We consider the following survival analysis model. Suppose there are $n$ subjects, each with $p$ covariates, denoted by $x_i=(x_{i1},\cdots,x_{ip})$, for  $i=1,2,\ldots n$. For each subject, there is also a pair of variables $(y_i,\delta_i)$, where $y_i$ denotes the time when the event happens, and $\delta_i$ is an indicator representing whether the subject fails ($\delta_i=1$) or loses to follow ($\delta_i=0$). We are interested in performing estimation and inference for the coefficients $\beta$. We are also interested in estimating the survival curve of a new subject based on its covariates and the training dataset.

In survival analysis, the Cox proportional hazards model (proposed in \cite{Cox1}, \cite{Cox2}) is a widely used model. In this model, the hazard function $\lambda(t|x_i)$ is modeled as
\begin{equation}
    \lambda(t|x_i)=\lambda_0(t)\exp(\sum_{j=1}^px_{ij}\beta_j),
\end{equation}
where $\beta=(\beta_1,\cdots,\beta_p)$ is the parameter, and the baseline hazard rate $\lambda_0(t)$ can be modeled flexibly.

For Cox proportional hazards model, in order to perform estimation and inference on the parameter $\beta$, we estimate $\beta$ that maximizes the partial likelihood
\begin{equation}
L(\beta)=\prod_{\delta_i=1}L_i(\beta)=\prod_{\delta_i=1}\frac{\lambda(y_i|x_i)}{\sum_{i':y_{i'}\geq y_i}\lambda(y_i|x_j)}=\prod_{\delta_i=1}\frac{\exp(\sum_{j=1}^p x_{ij} \beta_j)}{\sum_{i':y_{i'}\geq y_i}\exp(\sum_{j=1}^p x_{kj}\beta_j)}.
\end{equation}

Variable selection problem for Cox proportional hazard model has also been widely studied. To name a few, in \cite{Tibshirani} the lasso method is proposed for variable selection; in \cite{Fan} the smoothly clipped absolute deviation is proposed; and in \cite{Zhang} an adaptive lasso method is also proposed.

\subsection{Marrying survival analysis with machine learning}
We consider the estimation  of the Cox proportional hazards model, using the standard partial likelihood approach.  We note that at each time that a death occurs, the process can actually be viewed as a classification problem: whether or not a certain subject died at the specific time. Motivated by this, we propose a ``stacking'' idea that treats the survival analysis problem within the framework of classification. Combined with the stacking idea, powerful methods for classification in machine learning, such as logistic regression, random forests, gradient boosting machines and neural networks can all be used to model survival data. 

For stacking with logistic regression, we perform estimation and inference on the parameters, and the procedure is shown to be approximately equivalent to the method of maximizing partial likelihood in Cox proportional hazards model. For stacking with a general machine learning algorithm (for classification), we can estimate the survival curve for each new subject, and the procedure can outperforms the  standard linear Cox model (especially when there are complicated effects, such as interactions) in terms of estimated survival curves.

\section{Main idea}

In this part, we explain the main idea of our stacking method, and show  it can can be used to perform estimation in survival analysis. 

\subsection{The stacking idea}
The ``sequential in time''
  construction of the partial likelihood
 suggests a  way of recasting the survival problem as a two-class
 classification problem.

 Recall  that
 associated with each uncensored observation --- that is, each observation for which $\delta_i=1$ ---
 is a set of observations that are ``at risk'' at that observation's failure time $y_i$; this
 set can be formally written as
 $R(i)=\{i': y_{i'} \geq y_i \}$.
 Furthermore, we define $|R(i)|$ to be the number of observations in the risk set at time $y_i$.

 The idea is as follows. Assume that the $i$th observation is uncensored.
 We
 create a binary response of length $|R(i)|$, $\tilde{y}(R(i)) \in \{0, 1\}^{|R(i)|}$. The
 element of $\tilde{y}(R(i))$ corresponding to the $i$th observation is set to $1$, and all other elements of $\tilde{y}$ are set to zero.
 We also construct a predictor matrix $\tilde{X}(R(i))$ of dimension $|R(i)| \times p$, where $p$ is the number of features,
 consisting of the covariates associated
 with each observation in $R(i)$.

 We repeat this process for each uncensored observation, and we stack the predictor matrices and binary vectors obtained,
 in order to obtain a predictor matrix of dimension
 $\left( \sum_{i: \delta_i=1} |R(i)|\right) \times p$, and a binary response vector of length  $\sum_{i: \delta_i=1} |R(i)|$.
 Furthermore, we append to the  predictor matrix $\sum_{i=1}^n \delta_i$ binary columns, each of which contains all $1$'s for the elements
 corresponding to a particular risk set, and all $0$'s in the remaining elements.
 These two additional binary columns effectively allow a separate intercept corresponding to each risk set.

\bigskip
\begin{algorithm}
\caption{ Algorithm for Survival Data Stacking}
\label{alg1}
\begin{description}
\item For each risk set $R(i)$, create a binary response $\tilde{y}(R(i))$, and create a predictor matrix $\tilde{X}(R(i))$;
\item Stack predictor matrices $\tilde{X}(R(i))$ and binary vectors $\tilde{y}(R(i))$ together;
\item  Append binary columns to the predictor matrix, with $1$'s corresponding to risk set and $0$'s for the remaining elements.
\end{description}
\end{algorithm}

 We illustrate this construction through an example with three observations,
 $(x_1, y_1,\delta_1=1)$, $(x_2, y_2,\delta_2=0)$, and $(x_3, y_2, \delta_3=1)$. Furthermore,
 $y_1<y_2<y_3$,  and
 $x_i=(x_{i1},x_{i2})$ is  a predictor of length two.  The second observation is censored, while the other two are uncensored.
 The risk set corresponding to the first observation is $\{1,2,3\}$, and so we construct the predictor matrix
 $\tilde{X}(R(1)) =\begin{pmatrix}
 x_{11}&x_{12}   \\
 x_{21}&x_{22}  \\
 x_{31}&x_{32}  \\
 \end{pmatrix}$
 and the binary response vector $\tilde{y}(R(1))=\begin{pmatrix} 1 \\ 0 \\ 0 \end{pmatrix}$.
 We do not construct a risk set for the second observation, since it is censored.
 The risk set corresponding to the third observation is $\{3\}$, and so the predictor matrix is simply a row vector,
 $\tilde{X}(R(3))=\begin{pmatrix} x_{31} & x_{32}\end{pmatrix}$, and the response vector is simply a scalar, $\tilde{y}(R(3))=1$.

 We stack together $\tilde{y}(R(1))$ and $\tilde{y}(R(3))$ to obtain
 \begin{equation}
 \tilde{y}=\begin{pmatrix}
 1\\
 0\\
 0\\
 \hline
 1\\
 \end{pmatrix},\label{eq:ystar}
 \end{equation}
 and we stack together $\tilde{X}(R(1))$ and $\tilde{X}(R(3))$ to obtain
 \begin{equation}
 \tilde{X}=\begin{pmatrix}
 1&0&x_{11}&x_{12}   \\
 1& 0& x_{21}&x_{22}  \\
 1& 0& x_{31}&x_{32}  \\
 \hline
 0& 1& x_{31}&x_{32}  \\
 \end{pmatrix},
 \label{eq:xstar}
 \end{equation}
 where the first two columns of $\tilde{X}$ are indicator variables for whether the observations correspond to the risk set  for the
 first observation or the risk set for the third observation.
 In \eqref{eq:ystar} and \eqref{eq:xstar}, the horizontal lines indicate separation between the two risk sets.

 We now apply a binary classifier in order to predict $\tilde{y}$ using $\tilde{X}$: that is, to
 discriminate the individuals that failed at each failure time  from the other individuals in the risk set.
 In effect, we are modeling  the conditional probability of having an event at each failure time,
 having survived past the previous failure time.

 It turns out that if we apply a linear logistic regression classifier to the data $(\tilde{X},\tilde{y})$, 
 then we obtain parameter estimates  that are quite similar to those that result from fitting Cox's proportional hazards model.
 Table~\ref{tab:stacked} shows that on the time-to-publication data,
 there is good agreement between the two approaches.
 We note that when fitting a logistic regression model to $(\tilde{X}, \tilde{y})$, we do not include an intercept, as each
 risk set already has its own intercept associated with it: see the first two columns of \eqref{eq:xstar}.

According to Terry Therneau--- a leading  researcher  in survival analysis and the author of the widely used  R language package {\tt survival} -- in the context of
logistic regression, this idea is not new (personal communication).  
It arises, for we example, in the discussion of the relationship between discrete and continuous proportional hazards models
(see e.g. \cite{Therneau}). However we had difficulty finding specific references to the idea in print,  believe that it should be more widely known to statisticians and other data scientists. After the first version of this article was written, a reader (Justin Max)  pointed us to the excellent article  \cite{Allison}, which discusses the discrete approach in some detail.

 In a sense, this stacking trick is a ``poor man's"  approach to the
 proportional hazards model --- it allows us to fit a
 model that can accommodate censoring using simple software for binary classification.

 \begin{table}[ht]
\centering
\begin{tabular}{rrrrr}
\hline
& \multicolumn{2}{c}{Coefficient} & \multicolumn{2}{c}{$p$-value} \\
 & Cox  & Logistic  & Cox  &  Logistic  \\
 \hline
\textR{positiveresult[Yes]} & 0.571 & 0.402 & 0.001 & 0.010 \\
  \textR{multiplecenter[Yes]} & -0.041 & -0.011 & 0.871 & 0.962 \\
 \textR{clinicalendpoint[Yes]} & 0.546 & 0.504 & 0.037 & 0.038 \\
\textR{samplesize} & 0.000 & 0.000 & 0.751 & 0.678 \\
  \textR{budget} & 0.004 & 0.004 & 0.075 & 0.099 \\
 \textR{impact} & 0.058 & 0.050 & 0.000 & 0.000 \\
  \hline
\end{tabular}
\caption{\em Results for Cox's proportional hazards model and the stacked logistic regression approach on the time-to-publication data.
For brevity of exposition, the \textR{mechanism} variable is omitted. \label{tab:stacked}}
\end{table}

 When carrying out this stacking approach, we can use any two-class classifier in place of logistic regression: examples include
 trees, random forests, or boosting. These would facilitate the discovery of interactions.
 Similarly,  time-dependent covariates can be handled
 by using the predictor $x(t)$ at the appropriate timepoint when constructing the stacked feature matrix
 $\tilde{X}$.   Not everything is rosy with this approach, however.  $\tilde{X}$ can grow very large: with $n$ observations and
 no censoring, it has $n(n+1)/2$ rows. To deal with this, one can subsample the censored observations  from each of the risk sets.

\subsection{Estimation of survival curves}
The stacking idea can also be used for estimation of survival curves. We first use stacking with linear regression to illustrate the ideas, and then show how we can combine stacking with general machine learning algorithms to produce estimated survival curves. 

First for illustration, we consider stacking with linear regression. We call each risk set when a death happens a ``stratum''. For each subject $i$ in the $q$th stratum, the linear regression problem (where $\alpha_q$ is the intercept for the $q$th stratum)
\begin{equation}
    y_i=\alpha_q+\sum_{k=1}^px_{ik}\beta_k
\end{equation}
is equivalent to 
the following
\begin{equation}
    y_i-\bar{y}_q=\sum_{k=1}^px_{ik}\beta_k,
    \label{eq:center}
\end{equation}
where $\bar{y}_k$ is the mean of the $k$th stratum
(here, we assume that the covariates $x_{ik}$ are centered; if not, we simply center the them for each stratum). 
\subsection{Application of general ML methods}

Motivated by the above equivalence, we now show  how we can combine stacking with general machine learning algorithms to carry out survival analysis.  
General ML algorithms do not facilitate the inclusion of  intercepts for each stratum, so we instead use the centering idea discussed above.

We start with  stacked data,  centered as in (\ref{eq:center}), and we train a model using a machine learning algorithm (such as a random forests,  gradient boosting or neural networks). Let the fitted output function be denoted by $\hat{f}$. Now we are given a new observation $x_{new}=(x_{new,1},\cdots,x_{new,p})$. Suppose the ordered death times are $t_1<t_2<\cdots<t_l$, and suppose stratum $q$ corresponds to death time $t_q$. We can compute the column mean $\bar{M}_q$ of the stacked matrix for $q$th stratum and mean $\hat{\alpha}_q$ of the subvector of $y$ for the $q$th stratum. Then the predicted conditional death probability for the new subject at time $t_q$ (conditional on the event that the subject lives up to time $t_{q-1}$) is 
\begin{equation}
    \hat{\alpha}_q+\hat{f}(x_{new}-\bar{M}_q).
\end{equation}
Therefore, the predicted survival curve for the new subject from time $t_{q-1}$ to time $t_q$ is 
\begin{equation}
    \hat{P}(\text{survival until } t_q|\text{survival until } t_{q-1})= 1-(\hat{\alpha}_q+\hat{f}(x_{new}-\bar{M}_q)).
\end{equation}
Hence we can produce an  estimated survival curve using a general machine learning algorithm.

Moreover, we can generate a confidence band for the survival curve using the well-known Greenwood's formula. Specifically, suppose the estimated survival probability at period $q$ is $\hat{S}_q$, then the Greenwood's formula gives approximate standard error for $\hat{S}_q$ as follows:
\begin{equation}
    \text{sd}\{\hat{S}_q\}\approx \hat{S}_q(\sum_{j\leq q}\frac{y_j}{n_j(n_j-y_j)})^{\frac{1}{2}},
\end{equation}
where $n_j$ is the size of the risk set at period $j$, and $y_j$ is the number of deaths during period $j$.

This construction allows us to use any supervised learning method  based on squared error loss to carry out survival analysis.
That is, we use regression of a 0-1 outcome as a two-class classifier.
The use of classification methods based on binomial deviance is trickier, as the centering would be more difficult to finesse.
\section{Approximate equivalence  between stacking and partial likelihood estimation.}

In this part, we show the close relationship between stacking using logistic regression and maximizing partial likelihood by both theoretical analysis and through numerical simulations.

\subsection{Theoretical analysis}

For the Cox proportional hazards model, when there is a death for subject $i$, the contribution of the event to the   log partial likelihood is
\begin{equation}
    \eta_i-\log(\sum_{j\in R(i)}\exp(\eta_j))
\end{equation}
where $\eta_j=\sum_{k=1}^p x_{jk}\beta_k$.

Now suppose that we treat the same event in a logistic regression model, then the contribution to the binomial  log-likelihood is
\begin{equation}
    \beta_{0i}+\eta_i-\sum_{j \in R(i)}\log(1+\exp(\beta_{0i}+\eta_j)).
\end{equation}

If we optimize over $\beta_{0i}$ (by setting the derivative with respect to $\beta_{0i}$ to $0$)  to obtain
\begin{equation}
    \sum_j\frac{\exp(\hat{\beta}_{0i}+\eta_j)}{1+\exp(\hat{\beta}_{0i}+\eta_j)}=1.
\end{equation}
If we use the approximation
\begin{eqnarray}
1+\exp(\beta_{0i}+\eta_j) \approx 1
\label{eqn:approx}
\end{eqnarray}
then we have
\begin{equation}
    \hat{\beta}_{0i}\approx -\log(\sum_j\exp(\eta_j)).
\end{equation}
Hence the binomial og-likelihood is approximately
\begin{equation}
    \eta_i-\log(\sum_{j\in R(i)}\exp(\eta_j))-1
\end{equation}
which is the same as that for partial likelihood, up to a constant.
The approximation (\ref{eqn:approx}) works best for the large risk sets and will err
the most for events that occur near the end of the time period.

Alternatively, one can arrive at this approximation using the Poisson approximation to the partial log-likelihood
given by \cite{Whitehead} .

\subsection{Simulation studies: comparison of estimates and survival curves}

In this section, we compare the performance of stacking using logistic regression with maximizing partial likelihood using Cox model by simulation. In our simulation studies, we assume that the data is generated according to the following exponential hazard model:
\begin{equation}
    \lambda(t|x)=\exp(\beta^Tx),
\end{equation}
where we take $\beta=(-0.35,-0.2,0,-0.4,0,0)$. The covariates $x_{k}$ ($1\leq k\leq p$) are marginally standard normal, and are weakly correlated (the correlation between $x_j$ and $x_k$ is taken to be $\rho_{jk}=0.2^{|j-k|}$). We sample $10$ data sets independently; for each data set, there are $n=200$ subjects. (The maximum time is taken to be $1.5$.)

We first compare the coefficients and p-values of both methods in Figure 1. Each point in the figure represents a coefficient estimate/p value (with $x$ coordinate corresponding to the Cox model, and y coordinate corresponding to stacking). As can be seen from the figure, the results are almost identical.
\begin{figure}
\centering
\includegraphics[width=\textwidth]{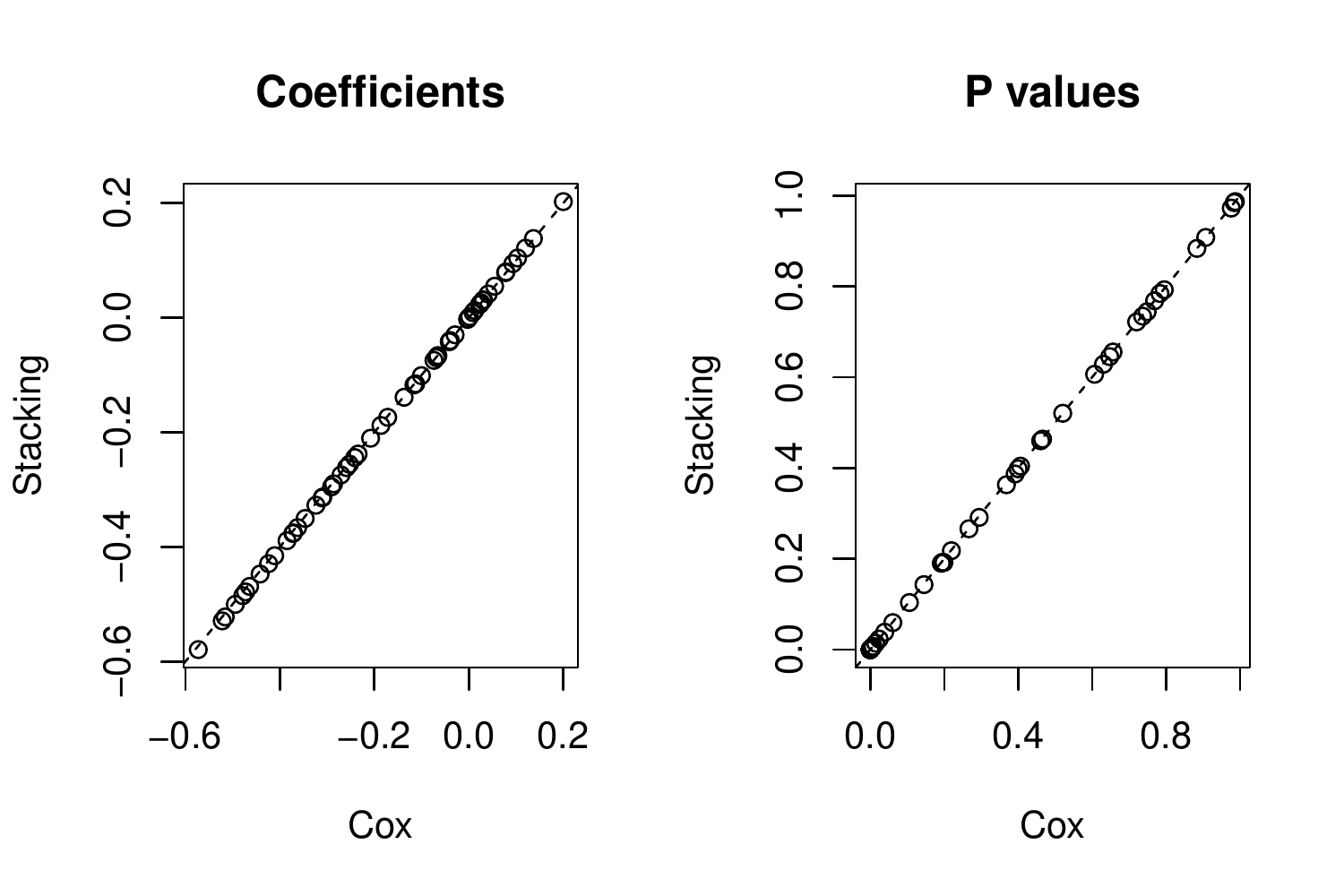}
\caption{\em Comparison of coefficients and p values: no penalty}
\end{figure}

Note that the parameter is sparse. In Figure 2 we show a similar comparison for $L_1$ penalized Cox model (the lasso method) and stacking with $L_1$ penalized logistic regression. The results are also very close.
\begin{figure}
\centering
\includegraphics[width=\textwidth]{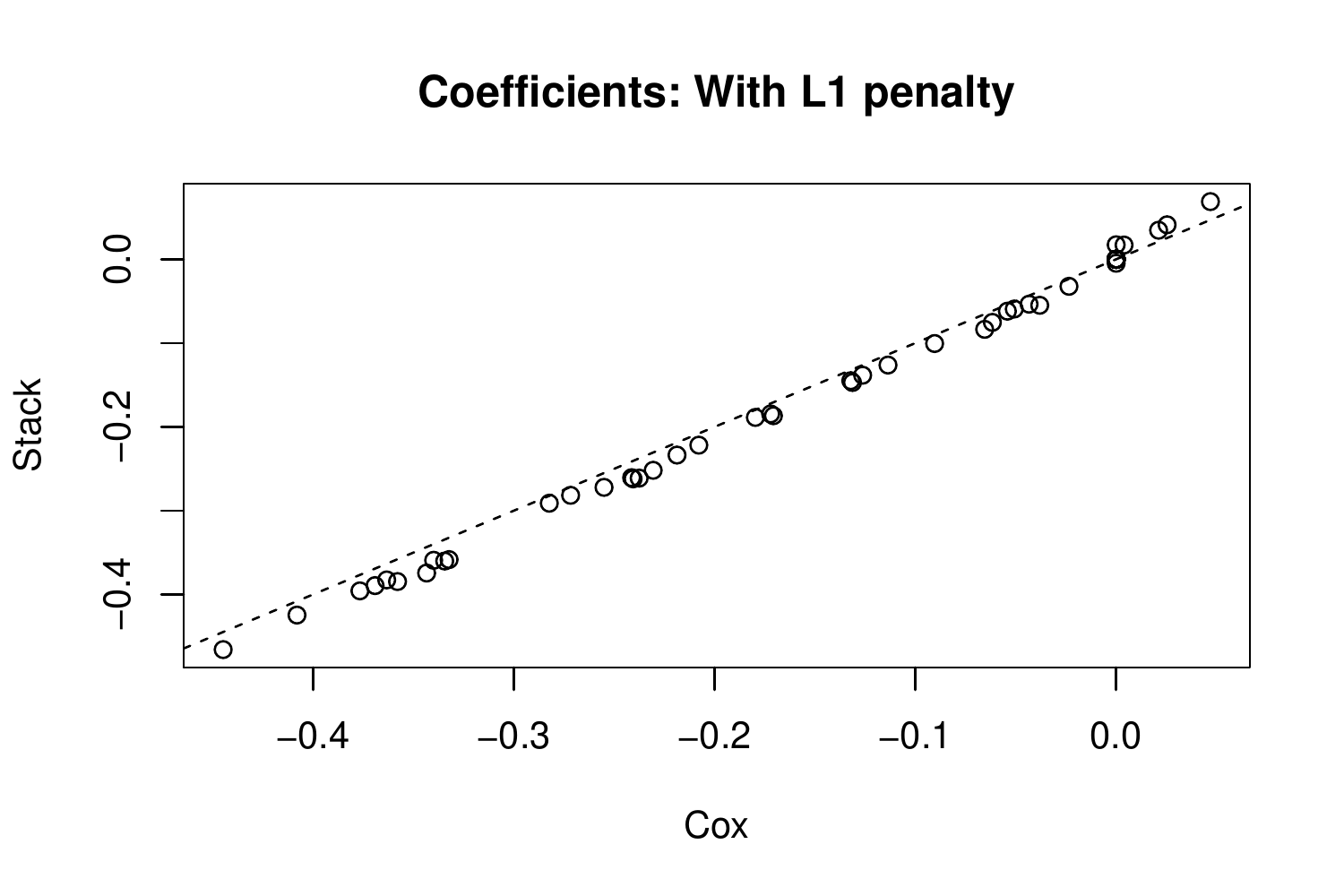}
\caption{\em Comparison of coefficients: $L_1$ penalty}
\end{figure}

Now we compare the estimated survival curve for a new subject. We take the same model, but with maximum time of $2$. We plot the estimated survival curves using Cox regression and stacking with least squares in Figure 3 (The true survival curve for generating the simulated data is also shown). From the figure, it can be seen that the estimated survival curves are quite similar. In the figure, the confidence band for Cox regression is generated from the R function \textR{coxph}, and that for stacking is produced using Greenwood's formula.
\begin{figure}
\centering
\includegraphics[width=\textwidth]{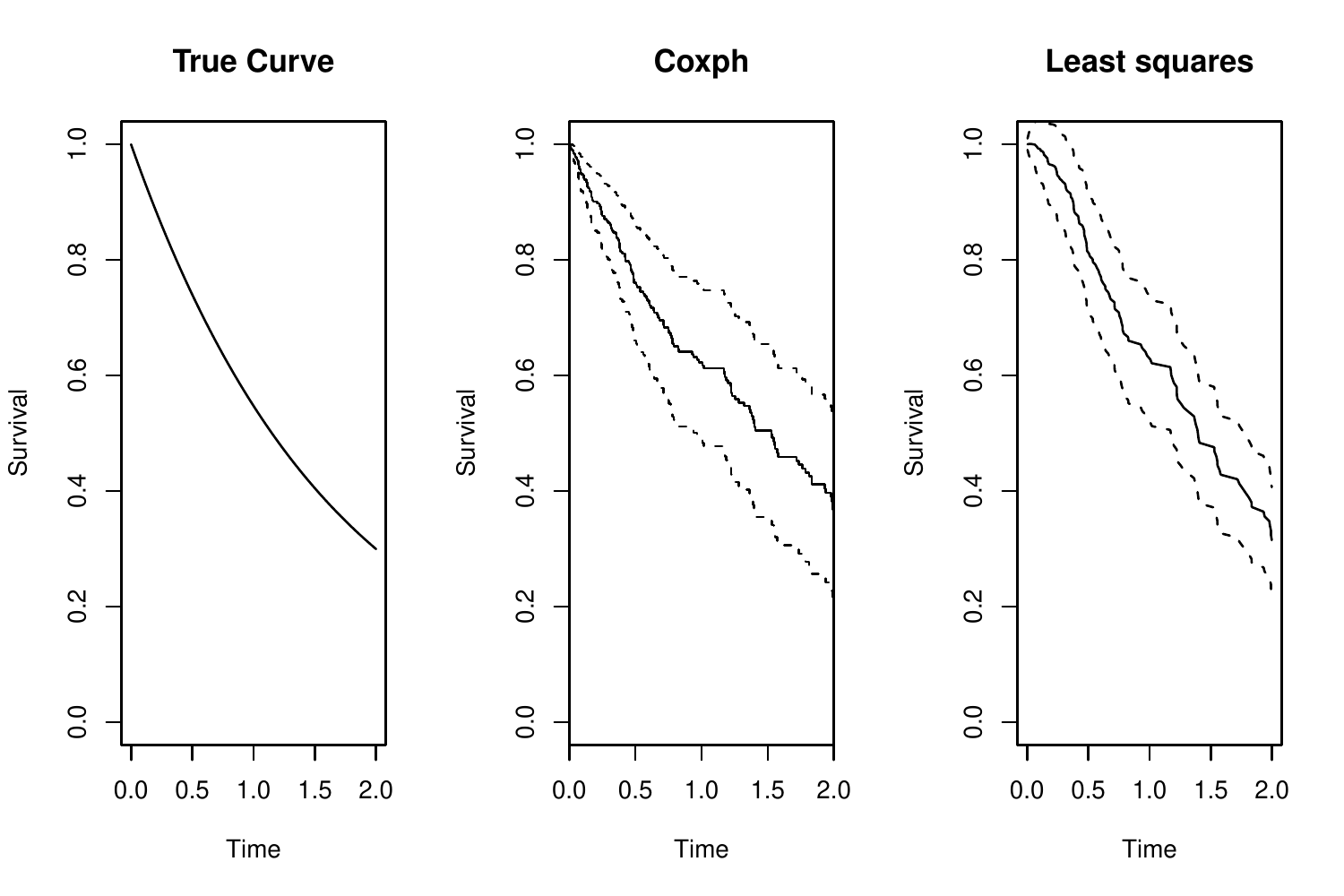}
\caption{\em Comparison of survival curves: Cox regression and stacking with least squares}
\end{figure}

\section{Extensions to other machine learning algorithms: simulation results}

As in the case for stacking with least squares, the stacking idea can be extended to other machine learning algorithms by plugging in a suitable learning algorithm $\hat{f}$. In this section, we present the simulation results for stacking with four  types of machine learning algorithms, namely, logistic regression, random forest, gradient boosting machine and neural networks for two types of models. The first model is suited for Cox regression, and the second model includes interaction effect (hence can be better estimated using machine learning algorithms such as a random forest). For both Model 1 and Model 2, we use a neural network with one hidden layer and two hidden units.

\subsection{Model 1}
In this part, we take the same model as before. The number of subjects is $n=200$. The estimated survival curves using  four methods are presented in Figure 4. It can be seen that in this case, the estimated survival curves are similar.
\begin{figure}
\centering
\includegraphics[width=\textwidth]{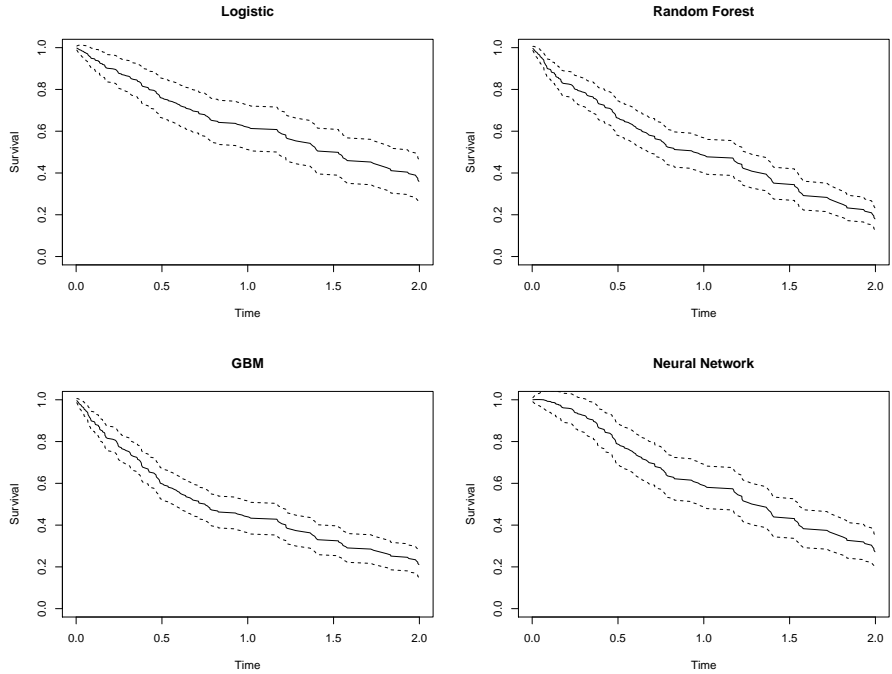}
\caption{\em Estimated survival curves for model 1}
\end{figure}

For comparison, we plot the estimated survival curves together in Figure 5. (We have also included a plot for the result using R function ``gbm'' with distribution ``coxph''.) It can be seen that for this model, the survival curves are quite similar. 
\begin{figure}
\centering
\includegraphics[width=\textwidth]{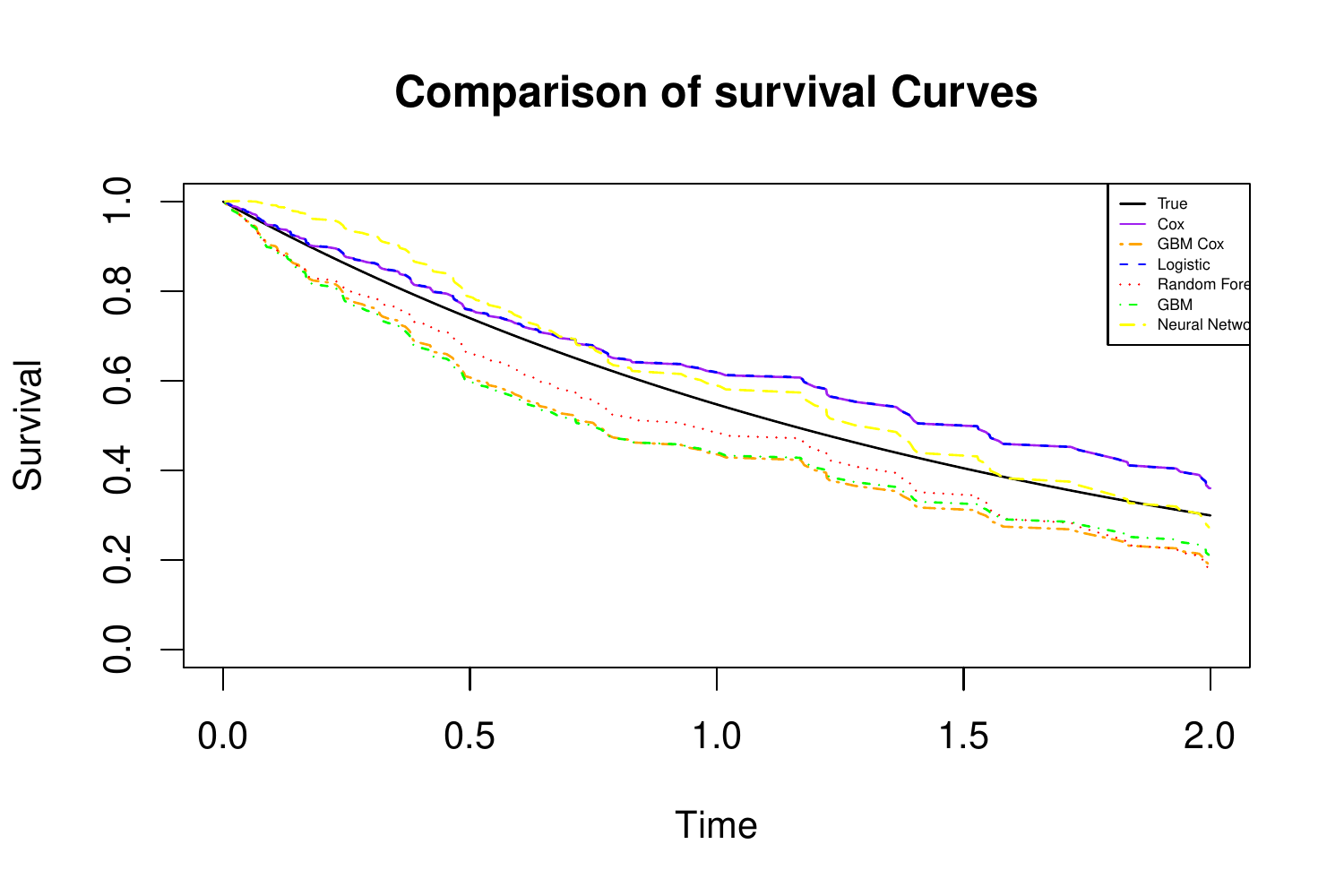}
\caption{\em Comparison of survival curves for model 1}
\end{figure}

\subsection{Model 2}
In this part, we study the performance of different methods on a model with non-linear and interaction effects. Specifically, the hazard rate function is of the form
\begin{equation}
    \lambda(t|x)=\exp(\beta_1x_5x_6+\beta_2x_1x_2+\beta_3x_3^2+\sum_{k=4}^6\beta_k x_k)
\end{equation}
with $\beta=(-0.35,0.2,0.45,0.6,0.8,0.01)$.
The number of subjects is also $n=200$. The corresponding survival curves are presented in Figure 6. (We have also included a plot for the result using R function ``gbm'' with distribution ``coxph''.) It can be seen that the survival curve generated from logistic regression is significantly different from those from other more complicated machine learning algorithms. For comparison, we also plot the estimated survival curves together in Figure 7. It can be seen that more flexible  machine learning algorithms such as random forests, gradient boosting and neural networks  give a better estimate of the true survival curve than the Cox models and logistic regression. This demonstrates the power of machine learning algorithms in survival analysis for models with high complexity (such as non-linear  and interaction effects).
\begin{figure}
\centering
\includegraphics[width=.9\textwidth]{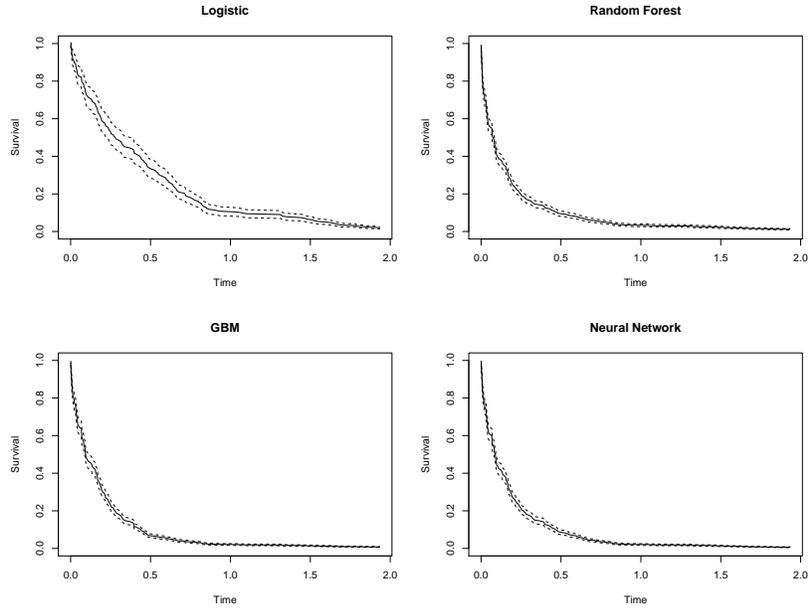}
\caption{\em Estimated survival curves for model 2}
\end{figure}
\begin{figure}
\centering
\includegraphics[width=.8\textwidth]{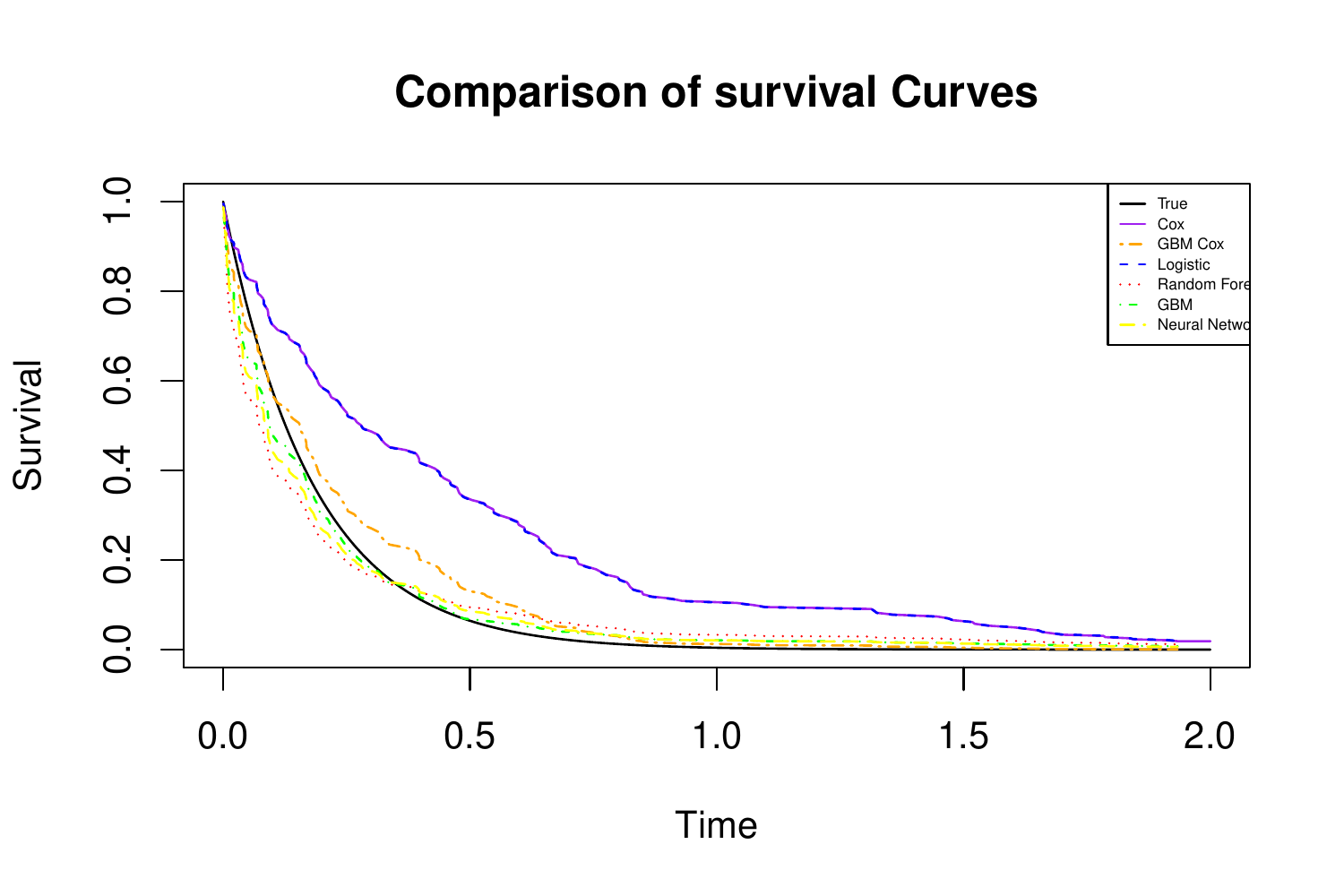}
\caption{\em Comparison of survival curves for model 2}
\end{figure}

\subsection{Evaluation using AUC}
Besides our comparison of survival curves, to further evaluate the performance of Cox regression and different stacking methods, we use Harrell's concordance (C) index to compute AUC for each method. Harrell's concordance index is as follows:
\begin{equation}
    C=\frac{\sum_{i,i'}I(y_i>y_{i'})I(\eta_{i'}>\eta_i)\delta_{i'}}{\sum_{i,i'}I(y_i>y_{i'})\delta_{i'}},
\end{equation}
where $\eta_i$ is the estimated risk for the $i$th subject. For stacking methods, to compare $\eta_i$ and $\eta_{i'}$, we compare the estimated survival curve at a time point in the middle as a substitute. Specifically, we randomly generate $20$ test data, use fitted models to predict their risk and compute the AUC values. The AUC values for Model 1 and Model 2 are displayed in Table 2.

We also include the AUC values when we compare the area under the estimated survival curve as a substitute of $\eta_i$ and $\eta_{i'}$. The AUC values are displayed in Table 3.

\begin{table}[ht]
\centering
\begin{tabular}{rrr}
\hline
 & Model 1  & Model 2 \\
 \hline
Cox regression & 0.743 & 0.727 \\
Stacking (logistic regression) & 0.743 & 0.727 \\
Stacking (random forest) & 0.686 & 0.658 \\
Stacking (GBM) & 0.743 & 0.781 \\
Stacking (neural network) & 0.777 & 0.620 \\
  \hline
\end{tabular}
\caption{\em AUC values for Model 1 and Model 2 using the first method \label{tab:stacked}}
\end{table}

\begin{table}[ht]
\centering
\begin{tabular}{rrr}
\hline
 & Model 1  & Model 2 \\
 \hline
Cox regression & 0.743 & 0.727 \\
Stacking (logistic regression) & 0.743 & 0.727 \\
Stacking (random forest) & 0.691 & 0.663 \\
Stacking (GBM) & 0.731 & 0.770 \\
Stacking (neural network) & 0.783 & 0.615 \\
  \hline
\end{tabular}
\caption{\em AUC values for Model 1 and Model 2 using the second method \label{tab:stacked}}
\end{table}

\section{Extension to time-dependent variables}
The stacking idea can also be extended to time-dependent variables. Specifically, when constructing the stacking data, we use the most recent values of the  covariates for each subject. For subject $i$, suppose the time-dependent covariates $x_{i,1},\cdots,x_{i,K}$ are measured at time $0=t_1<t_2<\cdots<t_K$ (where at time $t_K$ subject $i$ died or lost to follow). Then when constructing the stacking data for subject $i$ when an event happens at time $t$ with $t<t_K$, we use the covariate $x_{j,K}$ with $j$ the largest index such that $t_j\leq t$. 

To examine the stacking idea in time-varying covariate context, we perform simulation study using the built-in package in R function \textR{sim.survdata} to generate time-varying survival data. The comparison of estimated coefficients and p-values between stacking with logistic regression and Cox regression is shown in Figure 6. It can be seen that the results are very close, which supports the validity of our stacking idea in time-varying covariate context. 

\begin{figure}
\centering
\includegraphics[width=\textwidth]{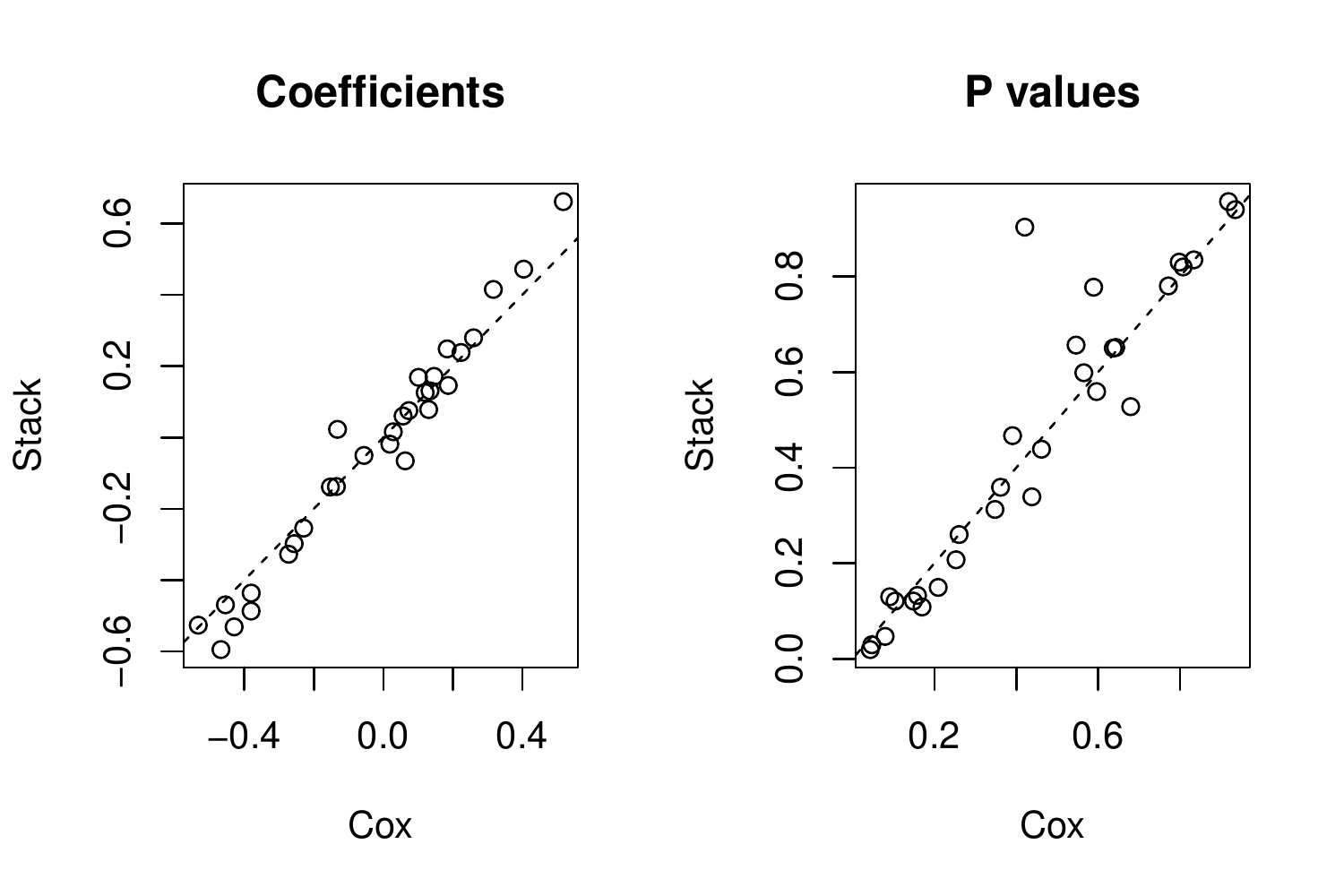}
\caption{\em Comparison of coefficients and p values: time varying covariates}
\end{figure}

\section{Discussion and conclusions}
In this work, we discuss a stacking idea  that  bridges survival analysis and machine learning. When we combine the stacking idea with logistic regression method, the estimation and testing results are very close the Cox proportional hazards model. Moreover, when combined with advanced machine learning algorithms for classification, our method can be used to tackle cases when the survival data is generated in a complicated manner (such as interaction effects). Therefore, the stacking method opens up the possibility of using machine learning ideas to perform survival analysis in a flexible and adaptive way. 

\medskip

{\bf Acknowledgements.}  We would like to thank Terry Therneau for  the argument in Section 3.1 and other helpful discussion.
 Robert Tibshirani was supported by NIH grant 5R01 EB001988-16 and NSF grant 19
DMS1208164.


\begin{thebibliography}{10}
\bibitem{Allison}
Allison, P.  ``Time Methods for the Analysis of Event Histories''.  Sociological Methodology, Vol. 13, (1982), pp. 61-98.
\bibitem{Cox1}
Cox, David R.``Regression models and life tables (with discussion)." Journal of the Royal Statistical Society 34.2 (1972): 187-220.
\bibitem{Cox2}
Cox, David R. ``Partial likelihood." Biometrika 62.2 (1975): 269-276.
\bibitem{Fan}
Fan, Jianqing, and Li, Runze. ``Variable selection for Cox's proportional hazards model and frailty model." The Annals of Statistics 30.1 (2002): 74-99.
\bibitem{Tibshirani} Tibshirani, Robert. The lasso method for variable selection in the Cox model." Statistics in Medicine 16.4 (1997): 385-395.
\bibitem{Therneau} 
Therneau, T. , and  Grambsch, P. (2001).
``Modeling Survival Data: Extending the Cox Model.'' Springer. New York.
\bibitem{Whitehead}
Whitehead, John. "Fitting Cox's regression model to survival data using GLIM." Journal of the Royal Statistical Society: Series C (Applied Statistics) 29.3 (1980): 268-275.
\bibitem{Zhang}
Zhang, Hao Helen, and Lu, Wenbin. ``Adaptive Lasso for Cox's proportional hazards model." Biometrika 94.3 (2007): 691-703.
\end{thebibliography}
\end{document}